\documentclass{mn2e}
\usepackage{natbib}
\usepackage{graphicx}
\usepackage{pslatex}

\begin{document}

\title[Stability of buoyant bubbles in clusters]{The stability of buoyant bubbles in the atmospheres of galaxy clusters}

\author[C.R. Kaiser, G. Pavlovski, E.C.D. Pope \& H. Fangohr]{Christian R. Kaiser$^1$\thanks{email: crk@soton.ac.uk}, Georgi Pavlovski$^1$, Edward C.D. Pope$^{1,2}$ \& Hans Fangohr$^2$\\ 
$^1$ School of Physics \& Astronomy, University of Southampton, Southampton SO17 1BJ\\
$^2$ School of Engineering  Sciences, University of Southampton, Southampton, SO17 1BJ
}

\maketitle

\begin{abstract}
The buoyant rise of hot plasma bubbles inflated by AGN outflows in galaxy clusters can heat the cluster gas and thereby compensate radiative energy losses of this material. Numerical simulations of this effect often show the complete disruption of the bubbles followed by the mixing of the bubble material with the surrounding cluster gas due to fluid instabilities on the bubble surface. This prediction is inconsistent with the observations of apparently coherent bubble structures in clusters. We derive a general description in the linear regime of the growth of instabilities on the surface between two fluids under the influence of a gravitational field, viscosity, surface tension provided by a magnetic field and relative motion of the two fluids with respect to each other. We demonstrate that Kelvin-Helmholtz instabilities are always suppressed, if the fluids are viscous. They are also suppressed in the inviscid case for fluids of very different mass densities. We show that the effects of shear viscosity as well as a magnetic fields in the cluster gas can prevent the growth of Rayleigh-Taylor instabilities on relevant scale lengths. R-T instabilities on pc-scales are suppressed even if the kinematic viscosity of the cluster gas is reduced by two orders of magnitude compared to the value given by Spitzer for a fully ionised, unmagnetised gas. Similarly, magnetic fields exceeding a few $\mu$G result in an effective surface tension preventing the disruption of bubbles. For more massive clusters, instabilities on the bubble surface grow faster. This may explain the absence of thermal gas in the north-west bubble observed in the Perseus cluster compared to the apparently more disrupted bubbles in the Virgo cluster. 
\end{abstract}


\section{Introduction}

Over the last few years numerous models have been proposed to explain the observed effects of non-gravitational heating of the gas in the centre of galaxy clusters. For a review of the observational evidence and the associated constraints on heating models see \citet{bmc02} and references therein. One of the most popular ideas involves the heating of the cluster gas by the outflows from AGN at the centres of clusters. The wide variety of outflows observed to originate in AGN has led to a number of suggestions for the detailed geometry of this model. These range from continuous, distributed heating throughout the cluster gas through the complete disruption of the outflow close to the centre of the cluster \citep{rb02,mb03} to episodic heating by collimated jets comparable to those of powerful radio galaxies \citep{rhb00,ba02,obb03}. 

The model geometry in which an episodic AGN inflates bubbles of relativistic plasma which subsequently detach from the cluster centre and rise buoyantly in the cluster atmosphere, has received most attention \citep{cbkbf00,qbb01,ssb01,bk01,mb03a,vbt04,rkf04} . There are a number of reasons for this concentrated effort. Firstly, although the complexity of the problem of modelling outflows from AGN in cluster atmospheres requires numerical simulations \citep[e.g.][]{jb01}, the geometry of individual bubbles rising passively due to buoyancy is comparatively simple to investigate. Secondly, although radio sources are common at the centres of cooling flow clusters \citep{jb90}, they rarely are powerful enough to inflate lobes on scales of several tens to hundreds of kpc. Finally, the observation of ellipsoidal depressions of the X-ray surface brightness profiles of clusters, sometimes associated with faint, low-frequency radio emission suggests this type of interaction of the AGN outflow with the cluster gas \citep[for an overview see][]{brm04}. 

A crucial factor in the heating model of buoyant bubbles is whether or not the bubbles are stable against break up by fluid instabilities on their surface. Clearly, in the case of complete disruption of the bubbles and mixing of the bubble material with the surrounding cluster gas, all energy stored in the bubbles can be dissipated. While this may be desirable from the point of view of efficiently heating the cluster gas, the apparently coherent bubble structures observed in clusters argue against complete mixing. Of course, it would be difficult in practice to find observational evidence in the cluster gas for bubbles after their disruption. We can therefore not rule out the possibility that clusters without observational signature of buoyant bubbles are still heated by this mechanism. For example, the large-scale radio structure of M87 in the Virgo cluster \citep{oek00} suggests the presence of buoyant bubbles \citep{cbkbf00,ck03a}, but there appears to be no corresponding depression in the X-ray surface brightness profile of the cluster \citep{ywm02}. 

In this paper we investigate with an analytical approximation the stability of the surface of buoyant bubbles rising in cluster atmospheres against disruption by fluid instabilities. We take into account the effects of shear viscosity of the cluster gas, which was recently suggested as a way to stabilise the bubblesÕ surface \citep{rkf04}. We also study the role of the cluster magnetic field by complementing and extending the analysis of \citet{dy03}. For simplicity in this analytical model, we assume a static cluster atmosphere in hydrostatic equilibrium in its gravitational potential well. There are no large-scale flows or turbulent motions present in the cluster gas. Such fluid flows may be caused by the continuing infall of gas, if the cluster is still in the process of forming, or by earlier outflows driven by the AGN. They may lead to a faster disruption of the buoyant bubbles, but their treatment is beyond the scope of analytical models. 

In Section \ref{instable} we discuss the relevant fluid instabilities. Section \ref{transport} contains a brief discussion of the relevant transport coefficients in ionised gases. We apply our estimates to the Virgo cluster in Section \ref{virgo} and compare our findings with the situation in the Perseus cluster in Section \ref{perseus}. We summarise our conclusions in Section \ref{conc}. The appendix contains a detailed derivation of the growth rates of fluid instabilities in the linear regime taking into account the effects of a gravitational field, surface tension, viscosity and relative motion between two fluids. 

\section{Instabilities on the bubble surface}
\label{instable}

Formally the surface of a buoyant bubble in the atmosphere of a cluster of galaxies represents a boundary between two fluids. In the presence of a gravitational field, this boundary will be subject to Rayleigh-Taylor (R-T) instabilities. Since the bubble is moving through the cluster atmosphere, Kelvin-Helmholtz (K-H) instabilities may also develop. Further complications arise from the possible presence of a magnetic field in the cluster gas and from the potentially relativistic equation of state of the material inside the bubbles. 

In the appendix we derive a general expression for the rate of growth, $n$, for instabilities of the boundary between two fluids as a function of the wavenumber of the instabilities, $k$, taking into account a gravitational field, viscosity, surface tension and motion of the two fluids relative to each other. 
Here we are mainly interested in the situation where the heavier fluid, i.e. the cluster gas, has a significantly higher mass density then the other fluid, i.e. the gas inside the bubbles. For this case we find in the appendix that the properties of the bubble material are unimportant in determining the growth rate of instabilities. Therefore all fluid quantities discussed in the following always refer to the heavy fluid, i.e. the cluster gas, unless stated otherwise.

The results in the appendix only apply to the non-relativistic case. However, in the simplest case where there is no surface tension, no viscosity and no motion of the two fluids relative to each other, the growth rate is given by the familiar form $n = \sqrt{gk}$, where $g$ is the gravitational acceleration at the fluid boundary. In the relativistic case this expression is modified to \citep{ah84}
\begin{equation}
n= \sqrt{\frac{gk \rho}{8p/c^2 + \rho}},
\end{equation}
where $\rho$ and $p$ are the rest mass density and the pressure of the cluster gas, respectively. For typical values of the particle number densities of $10^{-2}$\,cm$^{-3}$ and pressures of $10^{-11}$\,ergs\,cm$^{-3}$ in the cluster gas, the relativistic correction term $8p/c^2$, equivalent to the relativistic inertia of pressure, is at least five orders of magnitude smaller than $\rho$. In our regime we can thus safely neglect any corrections arising from a relativistic treatment of the gas in the bubbles.

In the appendix we also show that for fluids of very different mass densities K-H instabilities are suppressed completely on the boundary between the fluids. This result also holds for inviscid fluids. Therefore we will neglect K-H instabilities in the analysis below. 

For practical purposes the wavenumber, $k$, can be transformed into the corresponding length scale of the perturbation, $l= 2 \pi / k$. The growth rate $n$ gives the growth time $t_{\rm grow} = 1 / n$ which measures the time it takes the amplitude of a given perturbation to grow by a factor $e$. To first approximation the perturbation changes to the non-linear regime at this point and grows much faster afterwards. However, the instability takes a few $e$-folding times to fully develop \citep{dy03}.

\section{Transport coefficients in the cluster atmosphere}
\label{transport}

The temperature in clusters is a few to several keV, implying fully ionised gas. The viscosity of ionised, unmagnetised hydrogen is given by \citep{ls62}
\begin{equation}
\mu = 0.406 \frac{\sqrt{m_{\rm p}} \left( k_{\rm B} T \right)^{5/2}}{e^4 \ln \Lambda} \, \frac{{\rm g}}{{\rm s\,cm}}
\label{viscosity}
\end{equation}
where $m_{\rm p}$ is the proton mass. $k_{\rm B}$ is the Boltzmann constant, $T$ is the temperature of the gas and $e$ is the elementary charge. For our case of a pure hydrogen gas the Coulomb logarithm is
\begin{equation}
\Lambda = 24 \pi n_{\rm e} \left( \frac{8 \pi e^2 n_{\rm e}}{k_{\rm B} T} \right)^{-3/2},
\end{equation}
where $n_{\rm e}$ is the electron number density of the gas. The kinematic viscosity appearing in the equations of the appendix is then $\nu = \mu / \rho$. 

The importance of viscous processes in the transport of momentum in a fluid compared to inertial effects on typical length scales $L$ is given by the dimensionless Reynolds number, ${\rm Re} = u L / \nu$. $u$ is the typical velocity of the fluid on scales comparable to $L$. For cluster gas of a temperature $k_{\rm B} T \sim 3$\,keV and an electron density $n_{\rm e} \sim 0.1$\,cm$^{-3}$, we find ${\rm Re} \sim 100$ for transonic velocities on scales of roughly 1\,kpc. Obviously, ${\rm Re}$ will be even smaller for subsonic motion. We therefore conclude that viscosity can play an important role in the fluid dynamics of the cluster gas \citep{fsa03,rbb04}, if the kinematic viscosity is close to the value derived by Spitzer, i.e. if the cluster gas is fully ionised, but does not contain a magnetic field.

Thermal conduction is not directly relevant for the stability or otherwise of the surface of buoyant bubbles in clusters discussed in this paper. Nevertheless, it is interesting to note here that the transport of heat by conduction in the cluster gas is intimately linked with viscosity as both depend essentially on the same microscopic processes between gas particles. In the case of an unmagnetised plasma as described by \citet{ls62}, the thermal conductivity of the plasma is given by
\begin{equation}
\kappa = 4.5 \left( \frac{2}{\pi} \right)^{3/2} \frac{\left( k_{\rm B} T \right)^{5/2} k_{\rm B} }{\sqrt{m_{\rm e}} e^4 \ln \Lambda} \, \frac{{\rm ergs}}{{\rm s\,cm\,K}},
\label{conductivity}
\end{equation}
where $m_{\rm e}$ is the mass of an electron. The thermal diffusivity of the plasma is $\chi = \kappa / \left( c_{\rm p} \rho \right)$, with 
\begin{equation}
c_{\rm p} = \frac{5 k_{\rm B}}{2 \mu _{\rm m} m_{\rm p}},
\end{equation}
the specific heat at constant pressure. $\mu _{\rm m}$ is the mean mass of a gas particle in units of $m_{\rm p}$ and for pure ionised hydrogen $\mu _{\rm m} \sim 0.6$. To compare the relative importance of viscosity and thermal conduction for the transport of energy in the gas we can form the dimensionless Prandtl number ${\rm Pr}= \nu / \chi$. Using equations (\ref{viscosity}) and (\ref{conductivity}) we find
\begin{equation}
{\rm Pr} \sim 0.4 \sqrt{\frac{m_{\rm e}}{m_{\rm p}}} \sim 0.01.
\end{equation}
To first approximation Pr does not depend on the exact hydrodynamical conditions within the plasma, at least if the plasma is unmagnetised. This result means that if viscous effects are important in the cluster gas, then thermal conduction of heat will also play a significant role in the energy budget of the cluster gas. The latter effect has been invoked by a number of authors \citep{vsf02,fvm02,zn02,mb03,vf04} to explain the observed lack of cold gas in the centres of cooling flow clusters. If, for example, the viscous dissipation of the energy stored in sound waves excited by the outflows from AGN contributes to the heating of cluster gas \citep{fsa03,rbb04}, then thermal conduction should be important too. However, because Pr is small, this line of reasoning cannot be reversed. The gas in cluster atmospheres, in which thermal conduction is important, is not necessarily viscous.

In the case of a magnetic field in the plasma, the thermal conduction of heat and therefore also the viscous transport of momentum is suppressed perpendicular to the fieldlines. In a magnetic field tangled on scales much smaller than the typical size of the plasma this leads to a reduced value for the average thermal conductivity, $\nu$ \citep{cc98,nm01}. However, in the presence of turbulent motions in the cluster gas, anomalous heat transport may also occur \citep{clh03}, increasing the effective thermal conductivity of the gas. Similar effects are likely to affect the viscosity of the gas. It is not clear which description of thermal conductivity or viscous transport of momentum is applicable to the magnetised plasma in galaxy clusters \citep[for the issue of thermal conductivity see][]{vsf02,fvm02,vf04} and the unique conditions in individual clusters may well rule out a generic answer to this question. 

\section{Application to the Virgo cluster}
\label{virgo}

We now proceed to apply the results of the appendix and Section \ref{transport} to the hydrodynamical conditions derived from X-ray observations in galaxy clusters. In most clusters the buoyant bubbles are observed as depressions in the X-ray surface brightness of the cluster gas and are located close to the centres, $\sim 20$\,kpc, of the respective clusters \citep{brm04}. Therefore we need reliable estimates of the gas density and temperature in the cluster atmosphere down to very small radii. This requirement is aggravated by the need to determine the gravitational acceleration of the cluster gas from the spatial derivatives of these properties. 

\subsection{Observational constraints}
\label{obs}

The high-resolution X-ray maps of the nearby Virgo cluster produced by {\sc chandra} and {\sc xmm-newton} allow the determination of the gas properties down to a few kpc from the cluster centre \citep[][hereafter G04]{mbf02,ywm02,gmp04}. G04 obtain a good fit to the electron density distribution using the fitting function
\begin{equation}
n_{\rm e} = \frac{n_1}{\left[ 1 + \left( r / r_1 \right)^2 \right]^{\alpha _1}} + \frac{n_2}{\left[ 1 + \left( r / r_2 \right) ^2 \right] ^{\alpha _2}},
\label{vdens}
\end{equation}
with $n_1 = 0.089$\,cm$^{-3}$, $n_2 = 0.019$\,cm$^{-3}$, $r_1=4.5$\,kpc, $r_2 = 21$\,kpc, $\alpha _1 = 1.5$ and $\alpha _2 = 0.7$. For the temperature distribution they use 
\begin{equation}
T =  T_0 -  T_1 \exp \left( - \frac{r^2}{2 r_2^2} \right),
\label{vtemp}
\end{equation}
with $k_{\rm B} T_0 = 2.4$\,keV and $k_{\rm B} T_1 = 0.78$\,keV.

Assuming hydrostatic equilibrium for the gas in the gravitational potential well defined by the dark matter halo of the Virgo cluster, it is also possible to determine the mass distribution (G04) and therefore the gravitational acceleration $g$ within this cluster. The result from this analysis agrees very well in the region of overlap with the mass estimates derived from the dynamics of the globular clusters and the stars of M87 \citep{rk01}. The well known properties of the Virgo cluster make it an ideal example for the analysis of the stability of a buoyant bubble.

The torus-shaped radio structure \citep{oek00} associated with the central galaxy M87 on scales of approximately 15\,kpc to the east of the cluster centre has been interpreted as a buoyant bubble caused by the AGN at its centre \citep{cbkbf00,ck03a}. The situation is less clear to the west of M87, but a similar, more disrupted buoyant structure may be present there. However, no corresponding depressions of the X-ray surface brightness have been observed \citep{ywm02}. The eastern radio structure resembles a torus, inclined towards our line of sight. This is fully consistent with the structure expected for a buoyant bubble deformed, but not yet fully disrupted, by fluid instabilities on scales comparable to the bubble size of $\sim 10$\,kpc \citep{zmp77,cbkbf00}. On smaller scales, below the resolution limit of current observations, fluid instabilities may already be fully developed and may have mixed the bubble material with the cluster gas. The relativistic, magnetised plasma initially injected into the bubbles gives rise to the synchrotron emission while the mixed-in thermal cluster gas is responsible for the X-ray emission observed in the same region. The elongated radio emission region connecting the buoyant bubbles with the centre of M87 may be caused by small pockets of relativistic plasma created and detached from the main bubbles by fluid instabilities operating on small scales. We would not expect to observe these features `trailing' behind the main bubbles if the bubbles were still fully intact. Thus the buoyant bubbles in Virgo represent a good test case for our study as they suggest that fluid instabilities on the surface of the currently rising buoyant bubbles have fully developed on pc-scales, but are only starting to disrupt the bubbles on kpc-scales. 

The disruption and mixing of the buoyant bubbles with the cluster gas may influence the observed radio emission. The bubbles were originally inflated by the central AGN with a magnetised relativistic plasma similar to that of the currently `active' lobes on scales of $\sim 3$\,kpc surrounding the observed jet and postulated counterjet. We would expect the currently observed synchrotron radio emission to originate from the initial bubble material and not from any mixed-in thermal cluster gas. If the mixed-in cluster gas was also strongly magnetised, then internal depolarisation should lead to very low levels of polarisation of the radio emission from the buoyant bubbles \citep{bb66}. The fact that the synchrotron radio emission from the radio structures interpreted here as buoyant bubbles is highly polarised at high frequency \citep[10.55\,GHz,][]{rmkw96}, seems to argue against the bubbles being well mixed in with the cluster gas. However, the interpretation of synchrotron radio polarisation data depends strongly on the exact assumed geometry of the emitting region \citep{rl84}. Furthermore, if the mixed-in cluster gas is not or only weakly magnetised, then it will not have a significant effect on the polarisation of the synchrotron emission from the bubble material. We will see below that our findings from the stability analysis of the bubble surface argue for at best a weak magnetic field in the cluster gas surrounding the bubbles.

\subsection{The effects of viscosity and a magnetic field}

The solutions of equation (\ref{sol}) clearly depend on the location within the cluster under consideration as all relevant cluster properties change as a function of distance $r$ from the cluster centre. This implies that instabilities on the bubble surface of different wavenumber $k$, or, equivalently, different length scale $l=2 \pi/k$, will be able to grow at different $r$. The buoyant bubbles are rising in the cluster atmosphere and so instabilities may start growing when a given bubble is close to the cluster centre, but may be `frozen' again as the bubble travels further out. Alternatively, instabilities of a certain scale length may be suppressed in the cluster centre, but start growing once the bubble rises to larger $r$.

\begin{figure}
\centerline{
\includegraphics[width=8.45cm]{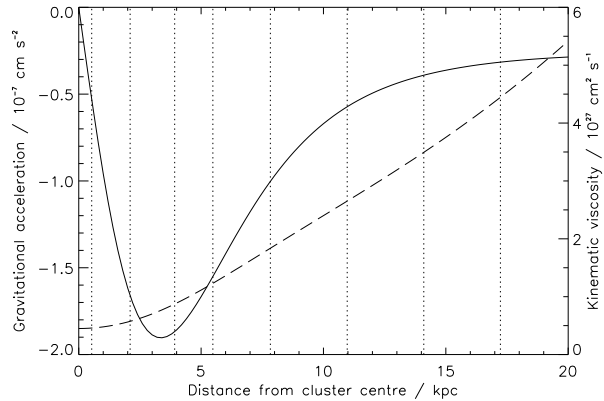}}
\caption{Gravitational acceleration (solid line) and kinematic viscosity (dashed line) of the gas in the Virgo cluster. Both quantities are derived from the analytical fit to the electron density and temperature profiles based on deprojected X-ray {\sc chandra} data by G04, equations (\ref{vdens}) and (\ref{vtemp}). The vertical dotted lines show the position of the data points used by G04. The kinematic viscosity is calculated for the case of a fully ionised, unmagnetised hydrogen plasma, equation (\ref{viscosity}).}
\label{gnu}
\end{figure}

\begin{figure}
\centerline{
\includegraphics[width=8.45cm]{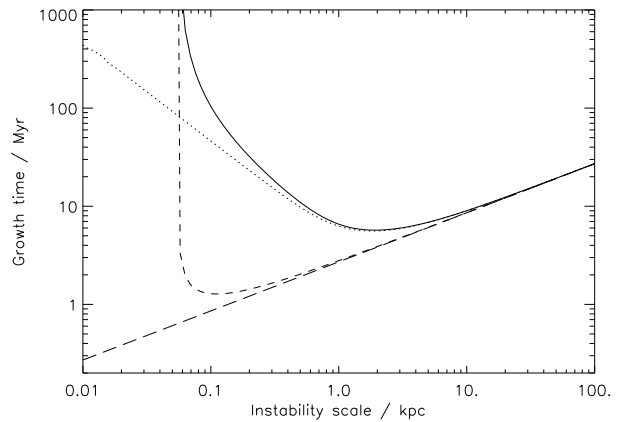}}
\caption{Growth time of fluid instabilities on the surface of a buoyant bubble at a distance of 10\,kpc from the centre of the Virgo cluster. Solid line: Real part of the fastest growing solution of equation (\ref{sol}) taking into account the viscosity of the cluster gas and a magnetic field of strength 1\,$\mu$G. Dotted line: Same as solid line, but neglecting the surface tension provided by the magnetic field. Short-dashed line: Same as solid line, but neglecting viscosity. This is the solution of equation (\ref{noviscos}). Long-dashed line: Growth time of instabilities without viscosity or magnetic field. Solution to equation (\ref{nonothing}).}
\label{modes}
\end{figure}

In the following we use the analytical fits to the electron density and gas temperature in the Virgo cluster presented by G04, equations (\ref{vdens}) and (\ref{vtemp}). Figure \ref{gnu} shows the resulting gravitational acceleration and the kinematic viscosity for an unmagnetised hydrogen plasma in hydrostatic equilibrium within the Virgo cluster. Using these results we now estimate the growth time, $t_{\rm grow}$, for fluid instabilities from equations (\ref{sol}), (\ref{noviscos}) and (\ref{nonothing}) as a function of their size $l$. Figure \ref{modes} illustrates the effects of the viscosity and a magnetic field on the growth time of the instabilities of various sizes. Note that these solutions only apply to a bubble located at a distance of 10\,kpc from the centre of the Virgo cluster. The situation changes considerably with the bubble position within the cluster (see below). 

In the absence of viscosity and surface tension from a magnetic field, instabilities grow on all size scales with the familiar proportionality of $t_{\rm grow} \propto \sqrt{l}$. Viscosity and magnetic fields only affect the growth of instabilities on small scales. The introduction of a magnetic field in the cluster gas results in a sharp cut-off at small sizes \citep[see also C61 and][]{dy03}. For illustration we used a magnetic field of 1\,$\mu$G in Figure \ref{modes}. Here and in the following we assume that the magnetic field is parallel to the bubble surface. This assumption is justified as any field components normal to the bubble surface will not influence the growth of instabilities in the linear regime (C61). Note that even such a weak magnetic field has a profound effect on the growth of the instabilities. Our assumption that the density of the cluster gas strongly exceeds that of the gas in the bubble leads to the disappearance of all quantities associated with the bubble material from the relevant expressions (see the appendix). The same holds for the magnetic field. Thus the magnetic field discussed here is the field in the cluster gas, not the field inside the bubble. The latter may be by far the stronger of the two, but, in our approximation, does not influence the growth of instabilities on the bubble surface. 

The cut-off due to the magnetic surface tension below which instabilities cannot grow is given by setting $t_{\rm grow} \rightarrow \infty$ or, equivalently, $n = 0$ in equation (\ref{noviscos}). Thus 
\begin{equation}
l_{\rm min} = \frac{B^2}{2 \rho g},
\label{lmin}
\end{equation}
and we would not expect a given bubble to disintegrate into units smaller than typically $l_{\rm min}$ \citep[see also][]{dy03}. The break up of the bubble into smaller bubbles of sizes only slightly larger than $l_{\rm min}$ happens rapidly. In fact, the minimum growth time is reached for an $l$ only slightly larger than $l_{\rm min}$. The structure of the bubble on large scales remains coherent for longer. If the bubble were observed from a distance with limited resolution, it would give the impression of being intact, while on small scales the observed large-scale structure is made up of many smaller units with a matrix of intermixed cluster gas in between them. Similar structures result when a coherent bubble of air is suddenly released under water \citep{zmp77}.

In the case of a viscous gas without a magnetic field, the growth of instabilities on the bubble surface is significantly delayed on small scales. There is no cut-off as in the case of a magnetic field, but again the growth time reaches a minimum and no instabilities on any size scale can grow faster than this minimum time. Finally, in the case of a viscous gas containing a magnetic field, the sharp cut-off introduced by the surface tension provided by the magnetic field is smoothed out somewhat by the effects of viscosity. 

\subsection{Stability of buoyant bubbles}

Figure \ref{modes} compares the effects of viscosity and the magnetic field in the cluster gas on the stability of the surface of a buoyant bubble. However, it only provides us with a snapshot at a single time during the rise of the bubble. As the bubble rises, the conditions for the growth of instabilities change and so we cannot use Figure \ref{modes} to estimate the degree of disruption of the bubble at a given time.

In the following we make the simplifying assumption that the buoyant bubbles in the Virgo cluster rise at a constant velocity of 400\,km\,s$^{-1}$. \citet{cbkbf00} showed that their numerical simulations of the bubbles in the Virgo cluster are consistent with a rise velocity of order the Keplerian velocity, $v_{\rm K} = \sqrt{g r}$, where $r$ is the distance from the cluster centre. For the inner 20\,kpc of the Virgo cluster, $v_{\rm K}$ is almost constant at 400\,km\,s$^{-1}$. The position of a buoyant bubble starting its rise at $t=0$ and $r=r_{\rm i} =1$\,kpc is simply given by $r=v_{\rm K} t +r_{\rm i}$. We obtain a rough estimate for the wavenumbers of instabilities that are able to grow on the bubble surface by setting $t_{\rm grow} = t$ and solving equations (\ref{sol}), (\ref{noviscos}) and (\ref{nonothing}) for $k$. We then convert the wavenumber to the relevant size scale, $l$, of the instabilities. We will assume for simplicity that the strength of the magnetic field in the cluster gas does not change as a function of distance from the cluster centre \citep[but see][]{mgf04}.

In the case of an inviscid cluster gas without a magnetic field, equation (\ref{nonothing}), this prescription results in an upper limit, $l_{\rm up}$, for the size of instabilities that are able to develop. Instabilities with $l \le l_{\rm up}$ should be well developed in this case. Taking into account viscosity or a magnetic field in the cluster gas results in a minimum for the growth time as a function of wavenumber or instability scale size (see Figure \ref{modes}). Depending on the rise time of the bubble, instabilities between a lower limit $l_{\rm low}$ and an upper limit $l_{\rm up}$ can develop or the bubble is stable on all scales. Unless the rise time of the bubble is close to the minimum growth time, $l_{\rm up}$ is almost identical for all cases.

\begin{figure}
\centerline{
\includegraphics[width=8.45cm]{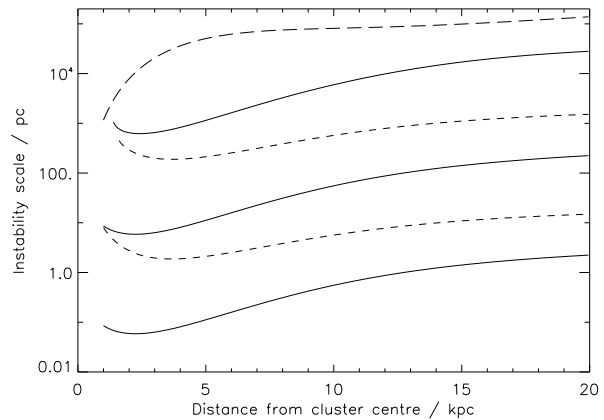}}
\caption{Limits for the scale size, $l$, of instabilities able to develop on the surface of a buoyant bubble in the Virgo cluster as a function of radius within the cluster atmosphere. Long-dashed line: Upper limit, $l_{\rm up}$, from equation (\ref{nonothing}) neglecting viscosity and magnetic fields. This upper limit also applies to good approximation to all other cases considered here (see text). Short-dashed line: Lower limit, $l_{\rm low}$, for a viscous, unmagnetised cluster gas with kinematic viscosities equal to the Spitzer value (top) and reduced by a factor 0.01 (bottom). Solid lines: Lower limit $l_{\rm low}$ for the case of an inviscid, but magnetised cluster with field strengths 0.1, 1 and 10\,$\mu$G from bottom to top. Instabilities on size scales between the lines giving lower limits, $l_{\rm low}$, and the long-dashed line at any fixed distance from the cluster centre, $r$, are able to develop within the rise time of the buoyant bubble currently located at $r$.}
\label{growth}
\end{figure}

Figure \ref{growth} shows the results for this rough estimate. In all cases the upper limit $l_{\rm up}$ starts to exceed the observed bubble size of about 10 kpc soon after the bubble starts rising. We therefore expect that the bubble should be deformed on scales comparable to its overall size. The observed, well-developed torus shape  of the eastern bubble and the apparently more advanced disruption on large scales of the western bubble are consistent with this prediction. For small instabilities the situation is more complicated. We pointed out in Section \ref{obs} that the absence of X-ray depressions in the region of the radio structures suggests the buoyant bubbles to be well mixed with the thermal cluster gas on pc-scales.  If the cluster gas has a viscosity as high as the value given by Spitzer's treatment, then no instabilities smaller than about 100\,pc can develop on the bubble surface at any stage during the rise of the bubble. Further out in the cluster atmosphere, viscous effects will suppress instabilities even up to 1\,kpc. For instabilities to be able to grow on pc-scales the kinematic viscosity has to be reduced by about two orders of magnitude. While such a significant suppression can be expected in the case of the related process of thermal conductivity due to the presence of magnetic fields in the cluster gas \citep{cc98}, it is currently not clear whether viscous effects are similarly reduced.

Turning to the effects of the magnetic surface tension, we find that even weak magnetic fields of 1$\mu$G already stabilise the bubble surface on scales down to about 10\,pc against R-T instabilities. The strength of the magnetic field in the cluster gas must be as low as about 0.1$\mu$G for instabilities on sub-pc scales to grow. Only for such low field strengths can we expect significant mixing of the bubble material with the thermal cluster gas. We will discuss our findings on viscosity and magnetic fields in more detail in Section \ref{conc}.

\section{Comparison with the Perseus cluster}
\label{perseus}

\subsection{Observational constraints}

The Perseus cluster is significantly further away from us than the Virgo cluster. Thus it is impossible to achieve the same accuracy in determining the gas properties and gravitational acceleration for the cluster regions inside 20\,kpc. However, we are mainly interested in studying the changes to the stability analysis for buoyant bubbles caused by a shallower gravitational potential. Furthermore, while we argued that the buoyant bubbles in the Virgo cluster are well mixed with the thermal cluster gas, the situation is quite different in the Perseus cluster. 

Two bubbles associated with the currently active AGN in the central galaxy NGC\,1275 are detected both as depressions in the X-ray surface brightness and in synchrotron radio emission \citep[e.g.][]{fse00}. There is at least one other depression in the X-ray surface brightness about 15\,kpc to the north-west of the cluster centre connected by a spur of low-frequency radio emission with the central radio source \citep{fcb02}. In the following we concentrate on this outer bubble, since it is located at a similar distance from the cluster centre as the buoyant bubbles in the Virgo cluster. The `hole' apparent in the X-ray maps is consistent with the responsible buoyant bubble being devoid of any thermal gas \citep{sfs02,sfa04}. It is therefore unlikely that small, pc-scale instabilities have mixed the bubble material with the cluster gas like for the bubbles in the Virgo cluster. On larger scales the bubble appears flattened in the rise direction and may also have been distorted into a torus shape similar to the eastern bubble in the Virgo cluster. This morphology indicates that the north-western bubble in the Perseus cluster may be in the process of deforming on scales comparable to the bubble size, $\sim 10$\,kpc, while its surface is stable on smaller scales. 

In order to derive the properties of the cluster gas and the gravitational acceleration of the Perseus cluster we use the mean deprojected electron density and temperature profiles presented in Fig. 21 of \citet{sfa04}. We fit the data points with the same expressions for the electron density and temperature we used for the Virgo cluster, equations (\ref{vdens}) and (\ref{vtemp}). However, there are only 3 data points inside a radius of 20\,kpc and so we simplify the density profile by setting $n_2=0$. We find a good fit for $n_1=0.07$\,cm$^{-3}$, $r_1 = 30$\,kpc and $\alpha _1 =0.85$. The corresponding fit for the temperature profile is $k_{\rm B}T_0 = 14.7$\,keV, $k_{\rm B} T_1 = 11.6$\,keV and $r_2 = 114$\,kpc. Note that the temperature of the gas in the Perseus cluster is still rising outwards at distances beyond 120\,kpc indicating a shallower gravitational potential than that of the Virgo cluster. The fits are based on only three data points inside the position of the buoyant bubble at about 20\,kpc from the centre. Clearly this rules out any attempt to derive results as accurate as for the Virgo cluster. However, we can still use these fits to illustrate systematic changes of our analysis for the given different properties of the two clusters.

\begin{figure}
\centerline{
\includegraphics[width=8.45cm]{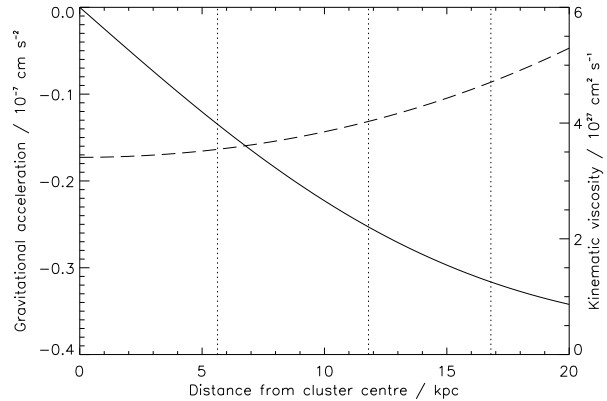}}
\caption{Gravitational acceleration (solid line) and kinematic viscosity (dashed line) of the gas in the Perseus cluster analogous to Figure \ref{gnu} for the Virgo cluster. The vertical dotted lines show the position of the data points inside 20\,kpc presented by \citet{sfa04}.}
\label{pgnu}
\end{figure}

From the density and temperature distributions in the Perseus cluster we derive the kinematic viscosity and gravitational acceleration, see Figure \ref{pgnu}. The values for both quantities can be compared with their values for the Virgo cluster, Figure \ref{gnu}. The kinematic viscosity is comparable in both clusters, but the gravitational acceleration is much lower at a given radius in the Perseus cluster. 

\subsection{Stability of buoyant bubbles}

\begin{figure}
\centerline{
\includegraphics[width=8.45cm]{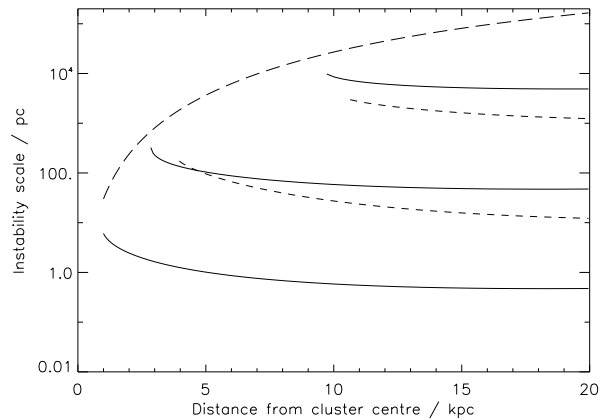}}
\caption{Limits for the scale size, $l$, of instabilities able to develop on the surface of a buoyant bubble in the Perseus cluster as a function of radius within the cluster atmosphere. Lines are as in Figure \ref{growth}, but calculated for the conditions in the Perseus cluster.}
\label{per_growth}
\end{figure}

Due to the monotonic behaviour of the gravitational potential inside 20\,kpc, the Keplerian velocity is monotonically rising. However, given the uncertainty for the value of the gravitational acceleration and for simplicity we make the same assumption as for the Virgo cluster that the rise velocity of any buoyant bubble is constant. We set it to 300\,km\,s$^{-1}$. 

Figure \ref{per_growth} shows the limits on the sizes of fluid instabilities that can grow on the surface of a buoyant bubble in the Perseus cluster. This is directly comparable to the situation in the Virgo cluster presented in Figure \ref{growth}. The upper size limit for the instabilities able to grow is lower in the Perseus cluster compared to the Virgo cluster. However, we would still expect well-developed instabilities on scales of several kpc at radii $\sim 20$\,kpc, consistent with the observations of the north-western bubble in the Perseus cluster. Viscous effects prevent the growth of instabilities below a size of about 1\,kpc if the kinematic viscosity is equal to that of an unmagnetised plasma. Even for a viscosity reduced by two orders of magnitude the bubble surface is stable down to several 10s of pc. The apparent absence of thermal material from the bubble on small spatial scales could therefore be explained by viscous effects. However, the surface tension provided by the cluster magnetic field would have the same effect for field strengths as low as a few tenths of $\mu$G. Mixing of the bubble material with the cluster gas on sub-pc scales is therefore unlikely within the Perseus cluster. This result contrasts with the situation in the Virgo cluster and is mainly caused by the smaller gravitational acceleration in Perseus. 

\section{Discussion and conclusions}
\label{conc}

We used the formalism of C61 to derive the linear growth rates of R-T and K-H instabilities on the boundary between two fluids. We demonstrated that the K-H instability is always suppressed if the fluids are viscous. Furthermore, the K-H instability is also negligible on all length scales for inviscid fluids, if the heavier fluid has a significantly larger mass density than the light fluid. We use our linear analysis to construct a simple prescription for the growth of fluid instabilities on the surface of light bubbles buoyantly rising in galaxy cluster atmospheres. Application of our prescription to the conditions derived from X-ray observations in the Virgo and Perseus clusters leads us to the following main conclusions. 

In the absence of magnetic fields and highly suppressed shear viscosity the surface of buoyant bubbles is highly susceptible to disruption by the R-T instability on all size scales up to the size of the bubbles. The complete mixing of bubbles with the thermal cluster gas will occur before the bubbles can rise beyond the innermost regions of the cluster. This behaviour is confirmed by numerical simulations down to the numerical resolution \citep[e.g.][]{bk01nat}. The mixing is complete on all length scales such that no trace of the initial geomtery of the bubble remains. 

Both shear viscosity and the presence of a magnetic field in the cluster gas can prevent the disruption of the bubbles on small scales. Viscous effects delay the growth of the instabilities while the magnetic field provides an effective surface tension. Mixing is prevented even if the kinematic viscosity is significantly suppressed by two orders of magnitude compared to the value given by Spitzer for a fully ionised gas. In the case of a magnetised cluster gas, a field strength of a few $\mu$G is sufficient to stabilise the bubble surface. 

An important factor in determining whether buoyant bubbles are disrupted in a given cluster atmosphere is the strength of the gravitational field. The gravitational acceleration sets the lower limit for fluid instabilities which are able to grow in the presence of a magnetic field via equation (\ref{lmin}). The effect of shear viscosity also depends on the gravitational acceleration. In general smaller instabilities can grow faster for larger gravitational acceleration. Thus the complete mixing of buoyant bubbles with the cluster gas should occur preferentially in cluster with a deeper gravitational potential well. 

In the case of the buoyant bubble observed in the Perseus cluster to the north-west of the cluster centre, the above considerations explain the coherent appearance of and the apparent absence of any thermal gas inside the bubble \citep{sfs02,sfa04}. The buoyant structures in the Virgo cluster observed in the radio, however, are not associated with depressions in the X-ray surface brightness and may therefore contain substantial amounts of thermal cluster gas. This observation suggests that the shear viscosity in the Virgo cluster must be highly suppressed to at least 1\% of the Spitzer value. Furthermore, the magnetic field strength in the cluster gas can be at most of order a few 0.1\,$\mu$G. 

We cannot measure the kinematic viscosity of the cluster gas directly and theoretical efforts to calculate transport coefficients in ionised, magnetised gases have so far concentrated on thermal conductivity. It is thus hard to assess whether a suppression of the kinematic viscosity in the Virgo cluster by at least two orders of magnitude is realistic. Even in the case of thermal conductivity estimates for the suppression factor range from 0.001 \citep{cc98} to 0.4 \citep{nm01}. The recent numerical simulations by \citep{rkf04} suggest that shear viscosity in the cluster gas leads to bubble morphologies and flow patterns consistent with observations. However, more effort is needed in the theoretical treatment of shear viscosity in magnetised plasmas to decide its importance in the cluster gas. 

We can estimate the strength of the magnetic field in the cluster from rotation measurements, RM, towards the central radio source in M87. \citet{oek90} find an average RM of about 1000\,rad\,m$^{-2}$ with peaks around 8000\,rad\,m$^{-2}$. If the Faraday screen responsible is distributed throughout the cluster gas, the implied strength of the magnetic field is 40\,$\mu$G. However, our line of sight to the inner radio lobes of M87 passes through the large-scale radio structure which will also act as a Faraday screen. It is possible that the entire observed RM originates in the outer radio emitting regions \citep{cbkbf00}. In this case, the magnetic field in the cluster gas itself is substantially lower. In general various methods employed in determining the strength of magnetic fields in clusters give different results \citep{ct02} and they may vary within individual clusters \citep{mgf04}. However, it is unlikely that the strength of the magnetic field in the vicinity of the buoyant bubbles in the Virgo cluster is much less than 1\,$\mu$G. This value may just be low enough to allow the mixing of the buoyant bubbles with the cluster gas on small length scales. 

In summary we find that buoyant bubbles should in general remain intact in cluster atmospheres. Exceptions like the bubbles in the Virgo cluster require a suppresion of the shear viscosity by more than two orders of magnitude compared to the value given by Spitzer. Furthermore, in such cases the strength of the cluster magnetic field must be below a few $\mu$G. It is therefore unlikely that we are failing to detect the remains of bubbles in many clusters of moderate mass because they have been disrupted and mixed in with the cluster gas. In the most massive systems with weak magnetic fields the disruption of bubbles may occur more easily. In these clusters the absence of depressions in their X-ray surface brightness profiles does not preclude the possibility that their atmospheres have been heated by buoyant bubbles in the past. 

This possibility is significantly enhanced if large-scale gas flows or turbulent motions are present in the cluster atmosphere. Both types of flows may be set up by gas infall in the earlier stages of cluster formation as well as the passage of bubbles caused by earlier activity cycles of the AGN through the cluster gas. As analytical models are unable to account for these additional effects, future numerical simulations of buoyant bubbles in clusters must address these issues. 

\section*{Acknowledgments}
The authors thank J.S. Sanders for providing us with an electronic version of the average data describing the Perseus atmosphere. We also thank the anonymous referee for helpful comments. ECDP thanks the Southampton Regional eScience Centre for support in the form of a studentship. CRK and GP thank PPARC for rolling grant support.

\def\newblock{\hskip .11em plus .33em minus .07em}

\bibliography{crk}
\bibliographystyle{mn2e}

\onecolumn
\appendix

\section{Instabilities at fluid boundaries}

Here we present a combined analysis taking into account the R-T and K-H instabilities as well as the effects of viscosity and surface tension at the boundary between two fluids superposed in  a gravitational field. We hereby extend the work of \citet[][hereafter C61]{sc61}, who neglected viscosity in his treatment of the K-H instability. A similar extension also, including a possible rotation of the two fluids, was attempted previously by \citet{bs84} and \citet{mb88}. However, we show below that their treatment of the problem is flawed. 

Consider a fluid of mass density $\rho$ and pressure $p$ in equilibrium in a gravitational field acting along the $z$-axis of the chosen coordinate system. The fluid has a coefficient of shear viscosity of $\mu$. For simplicity we assume that $\mu$ is a function of $z$ only. We assume that the velocity field in the fluid remains divergence free at all times, ${\bf \nabla \cdot v} =0$, which is equivalent to the fluid being incompressible. This condition is a good approximation as long as all motions remain subsonic \citep[e.g.][]{fs91b}. The fluid may stream uniformly in the direction of the $x$-axis at a velocity $U$. To allow for a situation where two fluids are superposed in the gravitational field, all quantities describing the fluid may change discontinuously at some $xy$-plane with coordinate $z_{\rm s}$. On this surface a possible surface tension $T_{\rm s}$ must be taken into account as well. 

Now consider a small disturbance of the fluid such that its density in the perturbed state is given by $\rho + \delta \rho$. The associated pressure perturbation is given by $\delta p$ and the perturbed velocities in the fluid are $U+u$ in the $x$-direction, $v$ in the $y$-direction and $w$ in the $z$-direction. We assume that the perturbation is small so that $u,v,w \ll U$. The surface at which discontinuities in the fluid properties occur is displaced in the $z$-direction by a $\delta z_{\rm s}$ due to the perturbation. Note here that the velocities $u$, $v$ and $w$ are perturbations and therefore are continuous at $z_{\rm s}$. The equations governing the subsequent evolution of the perturbation are then given by (check for equations 8 to 12 and 28 in chapter X and equations 1 to 6 in chapter XI of C61)
\begin{eqnarray}
\rho \frac{\partial u}{\partial t} + \rho U \frac{\partial u}{\partial x} + \rho w \frac{\partial U}{\partial z} & = & - \frac{\partial}{\partial x} \delta p + \left( \frac{\partial w}{\partial x} +\frac{\partial u}{\partial z} \right) \frac{{\rm d} \mu}{{\rm d} z} + \mu \nabla^2 u\\
\rho \frac{\partial v}{\partial t} + \rho U \frac{\partial v}{\partial x} & = & - \frac{\partial}{\partial y} \delta p + \left( \frac{\partial w}{\partial y} + \frac{\partial v}{\partial z} \right) \frac{{\rm d} \mu}{{\rm d} z} + \mu \nabla^2 v\\
\rho \frac{\partial w}{\partial t} + \rho U \frac{\partial w}{\partial x} & = & - \frac{\partial}{\partial z} \delta p + 2 \frac{\partial w}{\partial z} \frac{{\rm d} \mu}{{\rm d} z} + \mu \nabla^2 w - g \delta \rho + T_{\rm s} \left[ \left( \frac{\partial^2}{\partial x^2} + \frac{\partial^2}{\partial y^2} \right) \delta z_{\rm s} \right] \delta \left( z - z_{\rm s} \right)\\
\frac{\partial}{\partial t} \delta \rho + U \frac{\partial}{\partial x} \delta \rho & = & -w \frac{\partial \rho}{\partial z}\\
\frac{\partial}{\partial t} \delta z_{\rm s} + U_{\rm s} \frac{\partial}{\partial x} \delta z_{\rm s} & = & w_{\rm s}\\
\frac{\partial u}{\partial x} +\frac{\partial v}{\partial y} + \frac{\partial w}{\partial z} & = & 0,
\end{eqnarray}
where quantities with a subscript `s' are evaluated at the surface located at $z_{\rm s}$. 

We now consider perturbations of all fluid quantities of the form $f(z) \exp \left(i k_x x + i k_y y + nt \right)$, where $f(z)$ is an undetermined function which may be different for different quantities. Clearly only perturbations for which $n$ has a positive real part will grow in time. In the following we will derive expressions for $n$. 

Substituting this form for the perturbation into the above equations yields
\begin{eqnarray}
i k_x \delta p & = & - n \rho u + \mu \left( \frac{{\rm d}^2}{{\rm d} z^2} - k^2 \right) u + \left( i k_x w + \frac{{\rm d} u}{{\rm d} z} \right) \frac{{\rm d} \mu}{{\rm d} z} - i k_x \rho U u - \rho w \frac{{\rm d}U}{{\rm d} z} \label{e1}\\
i k_y \delta p & = & -n \rho v + \mu \left( \frac{{\rm d}^2}{{\rm d} z^2} - k^2 \right) v + \left( i k_y w + \frac{{\rm d} v}{{\rm d} z} \right) \frac{{\rm d} \mu}{{\rm d} z} - i k_x \rho U v \label{e2}\\
\frac{{\rm d}}{{\rm d} z} \delta p & = & -n \rho w + \mu \left( \frac{{\rm d}^2}{{\rm d} z^2} - k^2 \right) w + 2 \frac{{\rm d} \mu}{{\rm d} z} \frac{{\rm d} w}{{\rm d} z} - g \delta \rho -i k_x \rho U w -k^2 T_{\rm s} \delta z_{\rm s} \delta \left( z - z_{\rm s} \right) \label{e3}\\
n \delta \rho & = & -w \frac{{\rm d} \rho}{{\rm d} z} - i k_x U \delta \rho \label{e4}\\
\left( n + i k_x U_{\rm s} \right) \delta z_{\rm s} & = & w_{\rm s} \label{e5}\\
i k_x u + i k_y v & = &- \frac{{\rm d} w}{{\rm d} z} \label{e6},
\end{eqnarray}
where $k=\sqrt{k_x^2+k_y^2}$. By multiplying equation (\ref{e1}) by $-i k_x$, equation (\ref{e2}) by $-i k_y$, summing and making use of equation (\ref{e6}) we arrive at
\begin{equation}
k^2 \delta p = - n' \rho \frac{{\rm d} w}{{\rm d} z} + \mu \left(\frac{{\rm d}^2}{{\rm d} z^2} - k^2 \right)  \frac{{\rm d} w}{{\rm d} z} + \frac{{\rm d} \mu}{{\rm d} z} \left( \frac{{\rm d}^2}{{\rm d} z^2} + k^2 \right) w + i k_x \rho w \frac{{\rm d} U}{{\rm d} z},
\label{e12}
\end{equation}
where $n'=n+i k_x U$. Substituting into equation (\ref{e3}) from equation (\ref{e4}) for $\delta \rho$ and from equatuon (\ref{e5}) for $\delta z_{\rm s}$ yields
\begin{equation}
\frac{{\rm d}}{{\rm d} z} \delta p= - n' \rho w + \mu \left(\frac{{\rm d}^2}{{\rm d} z^2} - k^2 \right) w + 2 \frac{{\rm d} \mu}{{\rm d} z} \frac{{\rm d} w}{{\rm d} z} + \frac{g w}{n'} \frac{{\rm d} \rho}{{\rm d} z} - k^2 T_{\rm s} \frac{w_{\rm s}}{n'} \delta \left( z - z_{\rm s} \right).
\label{e22}
\end{equation}
We can now eliminate $\delta p$ between equations (\ref{e12}) and (\ref{e22}) to give
\begin{eqnarray}
\frac{{\rm d}}{{\rm d} z} \left\{ \left[ n' \rho - \mu \left(\frac{{\rm d}^2}{{\rm d} z^2} - k^2 \right) \right] \frac{{\rm d} w}{{\rm d} z}  - \frac{{\rm d} \mu}{{\rm d} z} \left( \frac{{\rm d}^2}{{\rm d} z^2} + k^2 \right) w - i k_x \rho w  \frac{{\rm d} U}{{\rm d} z} \right\} & = &\nonumber\\
= k^2 \left\{ \left[ -\frac{g}{n'} \frac{{\rm d} \rho}{{\rm d} z} + \frac{k^2}{n'} T_{\rm s} \delta \left( z - z_{\rm s} \right) \right] w_{\rm s} + \left[ \rho n' -\mu \left( \frac{{\rm d}^2}{{\rm d} z^2} - k^2 \right) \right] w - 2 \frac{{\rm d} \mu}{{\rm d} z} \frac{{\rm d} w}{{\rm d} z} \right\}. & &
\label{deq}
\end{eqnarray}

Equation (\ref{deq}) is a differential equation describing the possible solutions for the velocity of the perturbation in the $z$-direction, $w$. We are not interested in the exact form of $w$, but the solutions of equation (\ref{deq}) combined with its boundary conditions allow us to find expressions for $n$.

Here we are interested in cases where within a gravitational field a heavy fluid with mass density $\rho_2$ and viscosity $\mu_2$ is placed on top of a lighter fluid with mass density $\rho _1$ and $\mu_1$.  This situation can give rise to the R-T instability. We choose the location of the boundary between the two fluids such that $z_{\rm s} =0$, so that the heavy fluid is located at $z>0$ and the light fluid at $z<0$. For simplicity we assume that all fluid properties are uniform within the domain of each fluid and only change discontinuously at the boundary. This assumption also implies that both fluids move along the boundary at velocities $U_1$ and $U_2$, respectively. Since in general $U_1 \ne U_2$, the resulting shear at the boundary may cause K-H instabilities to grow. Note here that $n'$ also changes its value at the boundary. With these assumptions equation (\ref{deq}) reduces to
\begin{equation}
\frac{{\rm d}}{{\rm d} z} \left\{ \left[ \rho - \frac{\mu}{n'} \left( \frac{{\rm d}^2}{{\rm d} z^2} - k^2 \right) \right] \frac{{\rm d} w}{{\rm d} z} \right\} = k^2 \left[ \rho - \frac{\mu}{n'} \left( \frac{{\rm d}^2}{{\rm d} z^2} - k^2 \right)  \right] w
\end{equation}
inside the fluids, away from the boundary. By setting
\begin{equation}
q^2 = k^2 + \frac{n'}{\nu},
\end{equation}
where $\nu = \mu /\rho$ is the kinematic viscosity, the differential equation is simplified to
\begin{equation}
\left( \frac{{\rm d}^2}{{\rm d} z^2} - k^2 \right) \left(\frac{{\rm d}^2}{{\rm d} z^2} -  q^2 \right) w = 0.
\end{equation}
The solutions to this equation are given by linear combinations of the terms $\exp (\pm kz)$ and $\exp ( \pm q z )$. To avoid unphysical solutions of the form $w \rightarrow \infty$ for $z \rightarrow \pm \infty$, the solutions must have the form
\begin{eqnarray}
w_1 & = & n_1' \left( A_1 e^{kz} +B_1 e^{q_1 z} \right), \,{\rm for}\ z<0, \nonumber\\
w_2 & = & n_2' \left( A_2 e^{-kz} + B_2 e^{-q_2 z} \right), \,{\rm for}\ z>0.
\label{wsol}
\end{eqnarray}
Here and in the following subscripts denote quantities appropriate for the two individual fluids. The factors $n_1'$ and $n_2'$ are included in the solutions for convenience when applying the boundary conditions at the surface between the fluids further on.

C61 shows that four boundary conditions must hold at the surface between the two fluids. Equation (\ref{e5}) implies that $w/n'$ must be continuous at the boundary. Both \citet{bs84} and \citet{mb88} apply an incorrect boundary condition ($w_1 = w_2$) at this point. The same continuity at the boundary applies also to ${\rm d} w / {\rm d} z$ because of equation (\ref{e6}) combined with the continuity of $u$ and $v$. Note that if ${\rm d} w / {\rm d} z$ is not continuous or ill-defined at the boundary, then the last term of equation (\ref{deq}) implies that viscous stresses at this boundary grow without bounds until the continuity of $u$, $v$ and ${\rm d} w / {\rm d} z$ is established. C61 shows that the continuity of viscous stresses tangential to the boundary also imply that $\mu \left( {\rm d}^2/ {\rm d} z^2 + k^2 \right) w$ is continuous across the boundary. The final condition arises from integrating equation (\ref{deq}) across the boundary, i.e. from $0-\epsilon$ to $0+\epsilon$. From this integration we obtain
\begin{eqnarray}
\left. \left[ n_2' \rho_2 - \mu _2 \left( \frac{{\rm d}^2}{{\rm d} z^2} - k^2 \right) \right] \frac{{\rm d} w_2}{{\rm d} z} \right|_{z=+0} - \left. \left[ n_1' \rho_1 - \mu _1 \left( \frac{{\rm d}^2}{{\rm d} z^2} - k^2 \right) \right] \frac{{\rm d} w_1}{{\rm d} z} \right|_{z=-0} & = & \nonumber\\
k^2 \left\{ \left[ - g \left( \rho _2 - \rho _1 \right) + k^2 T_{\rm s} \right] \left. \left( \frac{w}{n'} \right) \right|_{z=0} - 2 \left( \mu _2 - \mu _1 \right) \left. \frac{{\rm d} w}{{\rm d} z} \right| _{z=0} \right\}.
\end{eqnarray}
Inserting the solutions given in equations (\ref{wsol}) in the boundary conditions results in four linear equations for $A_1$, $B_1$, $A_2$ and $B_2$. 
\begin{eqnarray}
A_1 + B_1 & = & A_2 + B_2\\
n_1' \left( k A_1 +q_1 B_1 \right) & = & - n_2' \left(k A_2 + q_2 B_2 \right)\\
\mu_1 n_1' \left[ 2 k^2 A_1 + \left( k^2 +q_1^2 \right) B_1 \right] & = & \mu_2 n_2' \left[ 2 k^2 A_2 + \left( k^2 + q_2^2 \right) B_2 \right]\\
- n_2'^2 \rho_2 \left(k A_2 +q_2 B_2 \right) + \mu_2 n_2' q_2 \left( q_2^2-k^2 \right) B_2 - n_1'^2 \rho_1 \left(k A_1 + q_1 B_1 \right) + \mu_1 n_1' q_1 \left(q_1^2-k^2 \right) B_1 & = & \nonumber\\
\frac{k^2}{2}\left\{ \left[ -g \left( \rho_2 - \rho_1 \right) +k^2 T_{\rm s} \right] \left( A_1 + B_1 + A_2 + B_2 \right) - 2 \left( \mu_2 - \mu_1 \right) \right.&& \nonumber\\
\left.\left[ n_1' \left( k A_1 +q_1 B_1 \right) - n_2' \left( k A_2 +q _2 B_2 \right) \right] \right\} & &
\end{eqnarray}
Here we have used 
\begin{equation}
\left.\left( \frac{w}{n'} \right)\right|_{z=0} = \left. \frac{1}{2} \left( \frac{w_1}{n_1'} + \frac{w_2}{n_2'} \right) \right|_{z=0}
\end{equation}
and
\begin{equation}
\left. \frac{{\rm d} w}{{\rm d} z} \right| _{z=0} = \left. \frac{1}{2} \left( \frac{{\rm d} w_1}{{\rm d} z} + \frac{{\rm d} w_2}{{\rm d} z} \right) \right|_{z=0}.
\end{equation}
in order to preserve the symmetry inherent in the boundary conditions. The set of linear equations can be conveniently written in matrix form,
\begin{equation}
\left( 
\begin{array}{cccc}
1 & 1 & -1 & -1\\
\\
n_1' k & n_1' q_1 & n_2' k & n_2' q_2\\
\\
2 k^2 \mu_1 n_1' & \mu_1 \left( q_1^2 + k^2 \right) n_1' & -2 k^2 \mu_2 n_2' & - \mu_2 \left( q_2^2 + k^2 \right) n_2'\\
\\
-k \alpha _1 n_1'^2 +\frac{k}{2} R - k n_1' C & \frac{k}{2} R - q_1 n_1' C & -k \alpha _2 n_2'^2 + \frac{k}{2} R + k n_2' C & \frac{k}{2} R + q_2 n_2' C
\end{array}
\right)
\left(
\begin{array}{c}
A_1\\
\\
B_1\\
\\
A_2\\
\\
B_2
\end{array}
\right)
=0,
\end{equation}
where we used the definitions of $q_1$ and $q_2$, divided all elements in the fourth row by $\left(\rho_1 + \rho _2 \right)$ and set
\begin{eqnarray}
\alpha _{1/2} & = & \frac{\rho_{1/2}}{\rho_1 + \rho _2}\\
C & = & k^2 \left( \alpha _1 \nu_1 - \alpha _2 \nu_2\right)\\
R & = & -k \left[ g \left( \alpha_1 - \alpha_2\right) + \frac{k^2 T_{\rm s}}{\rho_1 + \rho _2} \right].
\end{eqnarray}
The determinant of the matrix must vanish and this condition is equivalent to the required equation for $n$. We simplify the calculation of the determinant by dividing the elements of the first and second column by $n_1'$ as well as the elements of the third and fourth column by $n_2'$. We then subtract the first column off the second and the third column off the fourth. We then multiply the elements of the first column by $n_1'/n_2'$ and add this column to the third column. Finally, we multiply the elements in the third column by $n_2'$ and, after again using the definitions of $q_1$ and $q_2$, we obtain the reduced determinant
\begin{equation}
\left|
\begin{array}{ccc}
q_1 - k & k \left( n_1' +n_2' \right) & q_2 -k \\
\\
\alpha _1 n_1'  & 2 k^2 \left( \alpha _1 \nu_1 n_1' - \alpha _2 \nu_2 n_2' \right) & -\alpha_2 n_2'\\
\\
\alpha _1 n_1'+ C \left(1-\frac{q_1}{k} \right) & R + C \left( n_2' -n_1' \right) - \left( \alpha _1 n_1'^2 + \alpha_2 n_2'^2 \right) & \alpha_2 n_2' - C \left( 1- \frac{q_2}{k} \right)
\end{array}
\right|
=0.
\label{det}
\end{equation}
The equation resulting from expanding the above determinant is complicated. We do not reproduce it here, but note that it correctly reduces to equation (113) in chapter X of C61 for the case of $U_1 = U_2 =0$. 

In this paper we are specifically interested in cases where the mass density of the heavy fluid is much greater than that of the light fluid, i.e. $\rho_2 \gg \rho_1$. In this case, $\alpha_1 \sim 0$ and $\alpha_2 \sim 1$ and expanding the determinant in equation (\ref{det}) yields
\begin{equation}
q_2^4+2k^2 q_2^2-4k^3q_2 +k^4 -\frac{R}{\nu_2^2} = 0,
\label{sol}
\end{equation}
where we substituted $n_2' = \left( q_2^2-k^2 \right) \nu_2$. To arrive at equation (\ref{sol}) we have already divided out the trivial solutions $n_2'=0$, $q_1-k=0$ and $k^2 -q_2^2=0$, all of which only describe oscillations of the boundary with constant amplitudes. 

Several interesting conclusions can be drawn from our results. The properties of the light fluid are not important in determining the behaviour of instabilities at the boundary between the two fluids, since equation (\ref{sol}) does not depend on any quantities with a subscript `1'. Also, the velocities $U_1$ and $U_2$ do not explicitly appear in equation (\ref{det}). Let $N_1$ be an arbitrary solution of equation (\ref{det}) for $n_1'$ and $N_2$ the corresponding solution for $n_2'$. Then
\begin{eqnarray}
n & = & N_1 - i k_x U_1\nonumber\\
n & = & N_2 - i k_x U_2
\end{eqnarray}
give possible growth rates for instabilities. Since $N_1$ and $N_2$ must be independent of $U_1$ and $U_2$, these two velocities can only give rise to oscillations of the boundary with fixed amplitudes. In other words, in the linear regime K-H instabilities cannot grow on the boundary between two viscous fluids. This result is due to the last term in equation (\ref{deq}), which requires the continuity of ${\rm d} w/{\rm d} z$ for non-zero viscosities $\mu$. In the inviscid case K-H instabilities {\em can\/} grow, but only for wavenumbers exceeding (C61)
\begin{equation}
k_{\rm min} = \frac{g \left( \alpha _1 - \alpha _2 \right)}{\alpha _1 \alpha _2 \left(U_1 - U_2 \right)^2}.
\end{equation}
For our case of $\rho_1 \ll \rho_2$ we have $k_{\rm min} \rightarrow \infty$ and so the K-H instabilities are suppressed on all length scales.

The 4$^{\rm th}$-order polynomial in $q_2$ of equation (\ref{sol}) has four analytical, in general complex, solutions. The expressions for these solutions are very complicated and, unfortunately, cannot be turned into simple prescriptions for the existence of solutions for $n$ with positive real parts. However, in this paper we present plots of these solutions for the specific conditions investigated. 

If both fluids are inviscid, we note that $\nu_2 q_2 =0$ and we recover the solution presented in C61, i.e.
\begin{equation}
n = \sqrt{ gk \left( 1 - \frac{k^2 T_{\rm s}}{g \rho _2} \right)}.
\label{noviscos}
\end{equation}
In this form it is straightforward to see that surface tension leads to a maximum wavenumber $k_{\rm max} = \sqrt{g \rho_2 / T_{\rm s}}$, so that instabilities with $k > k_{\rm max}$ cannot grow (C61). The same $k_{\rm max}$ also applies in the viscous case. 

Finally, in the case of inviscid fluids without surface tension at the boundary we recover the well-known result
\begin{equation}
n = \sqrt{gk}.
\label{nonothing}
\end{equation}

\end{document}